\documentclass[a4paper,10pt,oneside,onecolumn,preprint,authoryear]{elsarticle}
\usepackage[utf8]{inputenc}
\usepackage{graphicx}
\usepackage{setspace}
\usepackage{multirow}
\usepackage{hyperref}
\def\DJ{\mbox{\raise0.3ex\hbox{-}\kern-0.4em D}}

\begin{document}

\title{Secular resonances with Ceres and Vesta}
\author[aob]{Georgios~Tsirvoulis\corref{cor1}}
\ead{gtsirvoulis@aob.rs}

\author[matf]{Bojan~Novakovi{\'c}}
\ead{bojan@matf.bg.ac.rs}

\address[aob]{Astronomical Observatory, Volgina 7, 11060 Belgrade 38, Serbia}
\address[matf]{Department of Astronomy, Faculty of Mathematics, University of Belgrade, Studentski trg 16, 11000 Belgrade, Serbia}

\cortext[cor1]{Corresponding author}

\begin{abstract}
In this work we explore dynamical perturbations induced by the massive asteroids Ceres and Vesta on main-belt asteroids through secular resonances. First we determine the location of the linear secular resonances with Ceres and Vesta in the main belt, using a purely numerical technique. Then we use a set of numerical simulations of fictitious asteroids to investigate the importance of these secular resonances in the orbital evolution of main-belt asteroids. We found, evaluating the magnitude of the perturbations in the proper elements of the test particles, that in some cases the strength of these secular resonances is comparable to that of known non-linear secular resonances with the giant planets.  Finally we explore the asteroid families that are crossed by the secular resonances we studied, and identified several cases where the latter seem to play an important role in their post-impact evolution.   
\end{abstract}

\begin{keyword}
 Asteroids, dynamics \sep Resonances, orbital \sep Asteroid, Ceres \sep Asteroid, Vesta
\end{keyword}

\maketitle
\section{Introduction}

For more than a century, a lot of studies have been done on the dynamical structure of the \textrm{Main Asteroid Belt}, a large concentration of asteroids with semi-major axes between those of Mars and Jupiter. Daniel Kirkwood proposed and then discovered the famous gaps in the distribution of main belt asteroids, which bear his name. The \textrm{``Kirkwood gaps''} are almost vacant ranges in the distribution of semi-major axes, corresponding to the locations of the strongest mean motion resonances with Jupiter, occurring when the ratio of the orbital motions of an asteroid and a planet (in this case Jupiter) can be expressed as the ratio of two small integers. Now we know that mean motion resonances with all the planets of the solar system exist throughout the main belt, rendering it a dynamically complex region. 

The gravitational interactions in the solar system also cause secular perturbations, which affect the orbits on long timescales. If the frequency, or a combination of frequencies, of the variations of the orbital elements of a small body becomes nearly commensurate to those of the planetary system, a secular resonance occurs, amplifying the effect of the perturbations. The importance of secular resonances has been pointed out already in the 19th century by \citet{Leverrier,Tisserand} and \citet{Charlier1,Charlier2}, who noticed a match between the $\nu_{\rm 6}$ secular resonance and the inner end of the main belt. A century later, thanks to the works of \citet{Froeschle1989,Morbidelli1991,Knezevic1991,Milani1992} and \citet{Michel1997} amongst others, we have a map of the locations of the most important secular resonances throughout the solar system. We thus have a clear picture of how the dynamical environment of the solar system, and consequently of the main belt, is shaped by the major planets.

Recently in \citet{Novakovic2015a} we have reported on the role of the linear secular resonance with (1)~Ceres on the post-impact orbital evolution of asteroids belonging to the (1726) Hoffmeister family. Contrary to previous belief, in which massive asteroids were only considered to be able to influence the orbits of smaller bodies by their mutual close encounters \citep{Nesvorny2002} and maybe low order mean-motion resonances \citep{Christou2012}, we have a concrete example that they can strongly affect the secular evolution of the orbits of the latter through secular resonances. Also, \citet{Li2016} have found that a secular resonance between two members of the Himalia Jovian satellite group can affect their orbital evolution, and \citet{Carruba2016} showed that secular resonances with Ceres tend to drive away asteroids in the orbital neighborhood of Ceres, giving more evidence that secular resonances in general can be important even if the perturbing body is relatively small. These results generate a number of questions regarding the potential role of such resonances in the dynamical evolution of the main asteroid belt in general. 

The scope of this work is to improve our general picture of the dynamical structure of the main belt, by studying the importance of the secular perturbations caused by the two most massive asteroids (1)~Ceres and (4)~Vesta.

\section{Methodology}
Here we describe the methods we followed to obtain a general picture of the effect of secular resonances with massive asteroids across the main belt. 

Our main analysis is based on numerical integrations of the orbits of test particles with initial conditions such that they cross the path of the resonance we are interested in, in order to evaluate the effect of the latter on their orbits.

To do so we first need to decide which secular resonances we should focus on.
Naturally, the first candidate for a perturbing body is (1)~Ceres, being the most massive asteroid, and having been proven to have a significant effect on the members of the Hoffmeister asteroid family through its linear nodal secular resonance\footnote{As in the usual nomenclature, $g$ \& $g_i$ stand for the proper frequency of the longitude of perihelion, and $s$ \& $s_i$ stand for the proper frequency of the longitude of the ascending node of the perturbed asteroid and the $i^{\rm {th}}$ perturbing body respectively. We will throughout this paper be using subscript ``c'' for Ceres and ``v'' for Vesta, while numbered subscripts refer to the fundamental frequencies of the Solar system.} $\nu_{\rm {1c}}=s-s_{\rm c}$ \citep{Novakovic2015a}. We are also considering secular resonances with the second largest asteroid, (4)~Vesta. Despite the fact that Vesta is less massive than Ceres, we wish to evaluate whether the perturbations arising from secular resonances with it are important for the dynamical evolution of asteroidal orbits, compared to other dynamical mechanisms.  We focus on the linear secular resonances with (1)Ceres and (4)Vesta, namely $\nu_{\rm {1c}}=s-s_{\rm c}$, $\nu_{\rm {c}}=g-g_{\rm c}$ and $\nu_{\rm {1v}}=s-s_{\rm v}$, $\nu_{\rm {v}}=g-g_{\rm v}$ as these resonances are expected to give rise to the strongest perturbations on the orbits of asteroids.

The first step of our study is to locate the path of each secular resonance we are interested in, across the main belt, in order to choose initial conditions for our test particles accordingly. While the locations of the secular resonances can easily be found analytically \citep{Knezevic1991}, giving a clear overall idea, we have found that the error of this approach for high eccentricities and inclinations is too high for the needs of our study, preventing us from accurately selecting the appropriate initial conditions that ensure interaction of the asteroids with the secular resonances. 

\begin{figure}[h!]
\begin{center}
\includegraphics[width=\columnwidth]{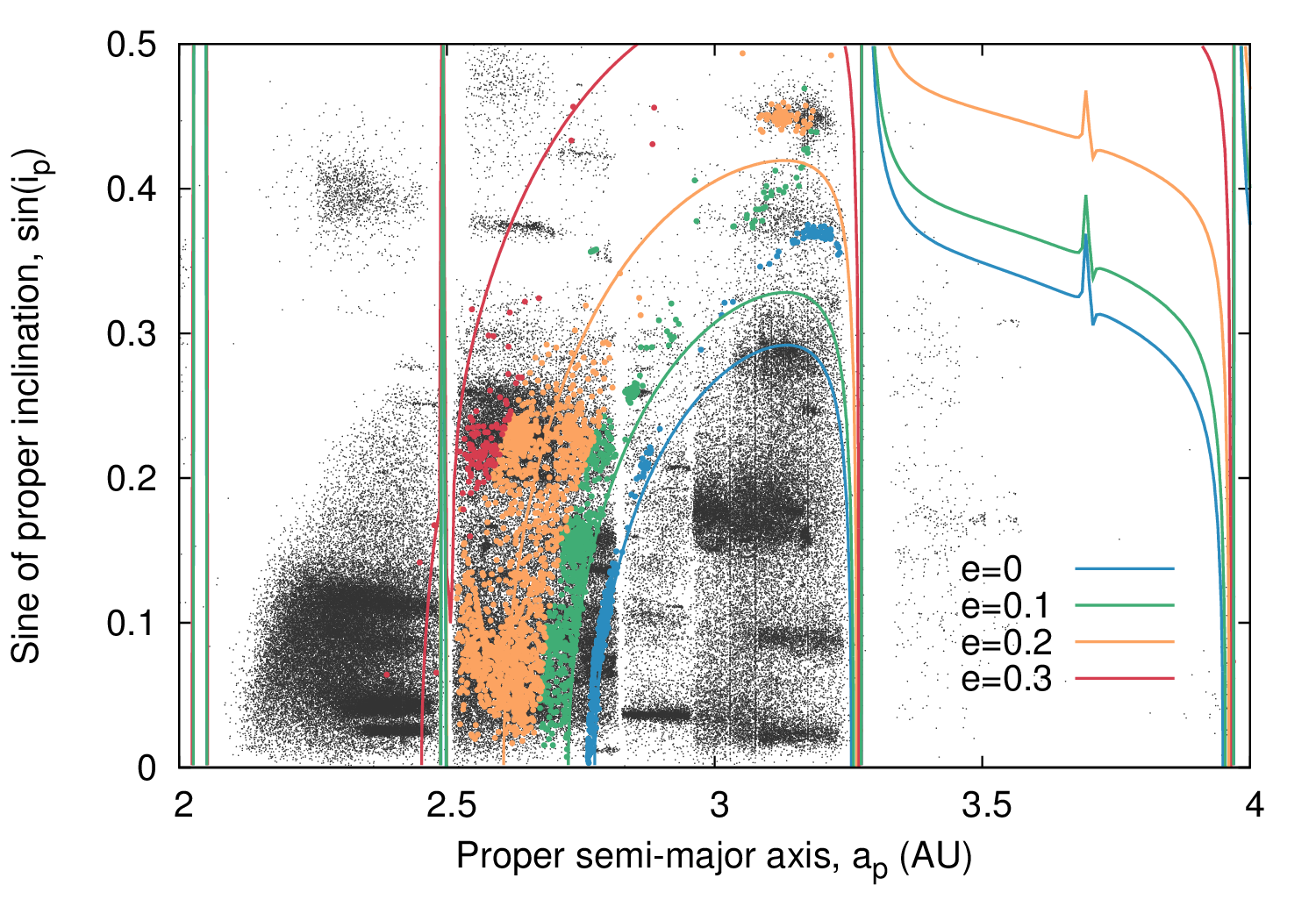} 
 \caption{The location of the $v_{\rm {1c}}$ secular resonance on the $(a_{\rm p},\sin{i_{\rm p}})$ plane. Solid lines represent the analytical solution for different values of the eccentricity (see legend). The colored dots show the resonant asteroids within $0.2''/{\rm {yr}}$, while the different colors correspond to values of eccentricity centered to those of the analytical solutions and spanning 0.05 in each direction}
   \label{fig:pathsv1c}
\end{center}
\end{figure}

We have decided thus to proceed with a different approach, based on the synthetic proper elements \citep{Knezevic2000}, and more specifically the proper frequencies, of the main belt asteroids, as released by the AstDyS service\footnote{available at: ${http://hamilton.dm.unipi.it/astdys2/}$}. From the catalog of proper elements we extract the proper frequencies of (1)~Ceres ($g_{\rm c}=54.07''/{\rm {yr}},\,s_{\rm c}=-59.17''/{\rm {yr}}$) and (4)~Vesta ($g_{\rm v}=36.87''/{\rm {yr}},\,s_{\rm v}=-39.59''/{\rm {yr}}$). Then, for a given secular resonance that we want to visualize, we select from the catalog those asteroids with proper frequencies that satisfy the corresponding resonant equation, within some margin corresponding to the strength of each resonance. To decide on the value of this margin we benefited from the analytical work of \citet{Knezevic1991}, where they use $2''/{\rm {yr}}$ for the most powerful secular resonance $\nu_{\rm 6}$, and $0.5''/{\rm {yr}}$ for weaker, fourth-degree resonances such as the $g+s-g_{\rm 6}-s_{\rm 6}$. Expecting that the secular resonances with massive asteroids should be relatively weak, we used $0.2''/{\rm {yr}}$ as a margin\footnote{The width of $0.2''/{\rm {yr}}$ used does not necessarily correpond to the width of the librating region of each resonance, which may vary across the main belt, but rather serves as a probe to easily visualize the paths of the resonances.} .  The asteroids with proper frequencies within these margins should lie along the path of the secular resonance in question. \autoref{fig:pathsv1c} shows an example of this approach for the secular resonance $\nu_{\rm {1c}}$, where the analytical solution is also plotted for comparison. Note the difference for high eccentricity and inclination between the two methods.

Having obtained the location of each secular resonance, we can decide on the parts of the main belt that should be studied. There is no strict rule for selecting initial conditions other than the proximity to the location of the secular resonance we examine in each case. We thus chose initial conditions in such a way, that a wide range of the proper elements of the main belt asteroids is sampled sufficiently for each case, as we will describe individually below. 

After selecting which parts we want to study, we proceed in the following way: We create groups of 20 fictitious particles with similar initial conditions and integrate their orbits for $50 \,{\rm {Myrs}}$, using the Orbit9  propagator\footnote{available within the $OrbFit$ package at: ${http://adams.dm.unipi.it/\sim orbmaint/orbfit/}$}, within two dynamical models: one including the four giant planets, from Jupiter to Neptune, and the massive asteroid relevant for each resonance as main perturbers\footnote{In these simulations we used values of $4.76\times10^{-10}$ and $1.3\times10^{-10} M_{\odot}$ for the masses of Ceres and Vesta, respectively \cite{Baer2011,Kuzmanoski2010}.}, and another one only with the four planets, which serves as a reference. Both dynamical models also incorporate the Yarkovsky effect as a secular drift in semi-major axis. This drift is expected to force the test particles to cross the resonance, causing the simulation that includes the massive asteroid as a perturber to reveal the resonant effect. We selected a value of ${{\rm d}a\over{\rm d}t}=4\cdot10^{-4}\,{\rm AU}\cdot {\rm {Myr}}^{-1}$ for the strength of the Yarkovsky induced drift, that may be considered as a typical reference value for asteroids of 1~km in diameter \citep{Vok2015}. This value allows for reasonably short integration times ($50 \,{\rm {Myrs}}$) while allowing enough time for the evolution of the test particles inside the resonance to investigate the respective perturbations. From the numerical integrations we obtain the time evolution of the asteroids' mean orbital elements. We then partition these in a running window manner in order to compute the time series of the synthetic proper elements \citep{Knezevic2000} for each asteroid. The comparison of the evolution of the test particles' proper orbital elements between the two dynamical models reveals the role of the secular resonances with the massive asteroids.

\section{Results}
\subsection{Secular resonances with Ceres}
\subsubsection{The $\nu_{\rm {1c}}$ resonance}
\begin{figure}[h!]
\begin{center}
\includegraphics[width=\columnwidth]{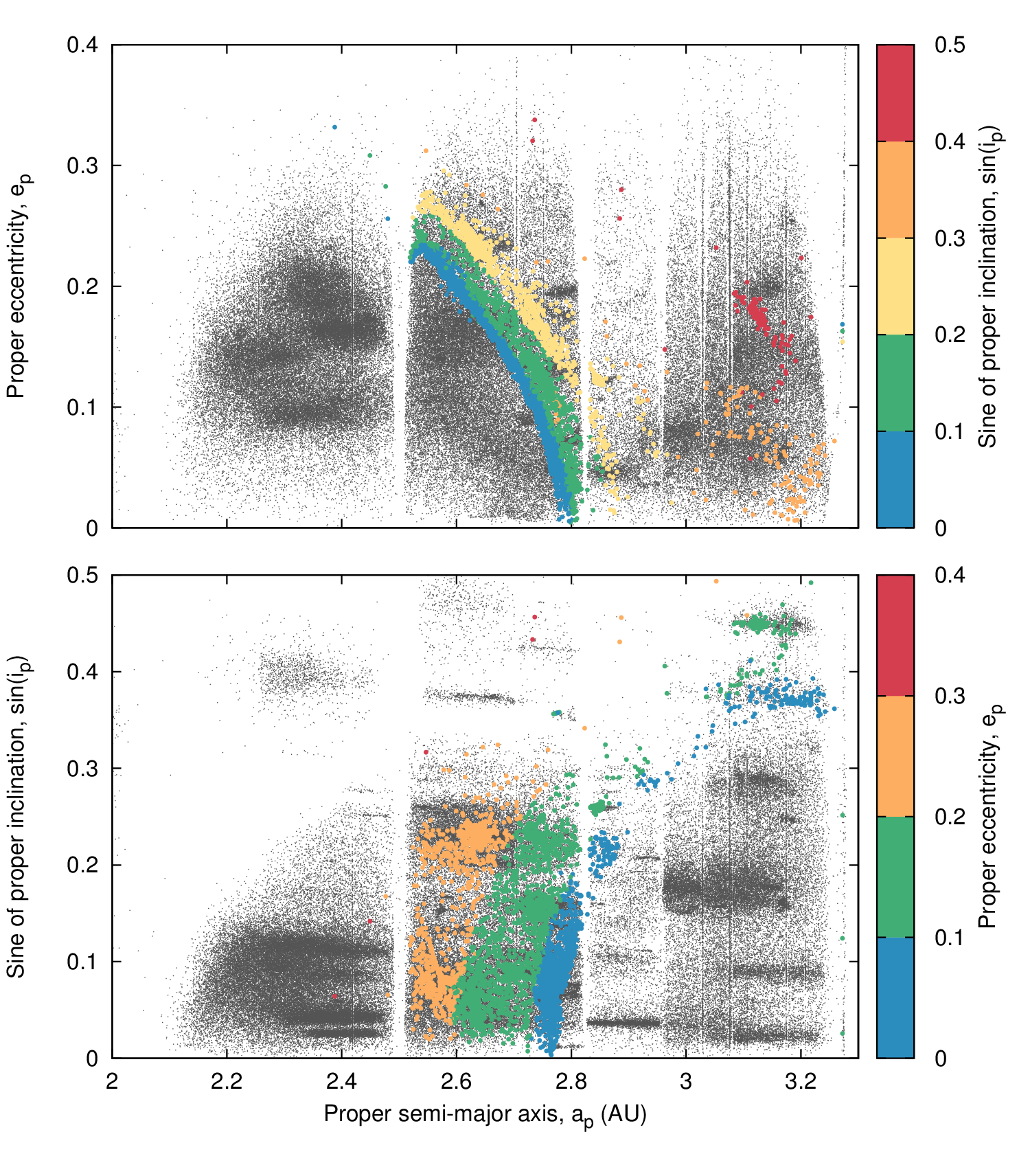} 
 \caption{The location of the $v_{\rm {1c}}$ secular resonance on the $(a_{\rm p},e_{\rm p})$ plane (top), and on the $(a_{\rm p},\sin{i_{\rm p}})$ plane (bottom). The gray dots represent all main belt asteroids, and the colored points the resonant ones for different inclinations (top panel) and eccentricities (bottom panel) according to the respective color codes given in the legend.}
   \label{fig:v1c}
\end{center}
\end{figure}

The first secular resonance we studied is the linear nodal secular resonance $\nu_{\rm {1c}}$. \autoref{fig:v1c} shows the proper semi-major axis versus the sine of proper inclination and the proper eccentricity  projections of the main asteroid belt. The resonant asteroids, the ones that satisfy the relation $|s-s_{\rm c}|\leq0.2''/{\rm {yr}}$, are highlighted, revealing the location of the resonance. Since the secular resonances are represented as surfaces in the three dimensional proper element space, we use a color code to grasp the third dimension when projecting on the plane.

We notice that the secular resonance crosses the middle $(2.5<a_{\rm p}<2.82\,{\rm {AU}})$ and outer $(2.82<a_{\rm p}<3.26\,{\rm {AU}})$ parts of the main belt. In the top panel of \autoref{fig:v1c}, we see that this resonance's projection on the $(a_{\rm p},e_{\rm p})$ plane appears as a wide strip that crosses the middle belt at an angle. This strip has a well defined lower boundary which corresponds to zero inclination, with the upper boundary being due to the gap in the distribution of asteroids at $\sin{i_{\rm p}}\sim0.3$. In the outer belt the resonant asteroids are less and more localized: two concentrations are found in the region $2.82<a_{\rm p}<2.9$ and another two at high inclinations past $3 \,{\rm {AU}}$, corresponding to asteroid families as will be discussed in the next section.   

\begin{figure}[h!]
\begin{center}
\includegraphics[width=\columnwidth]{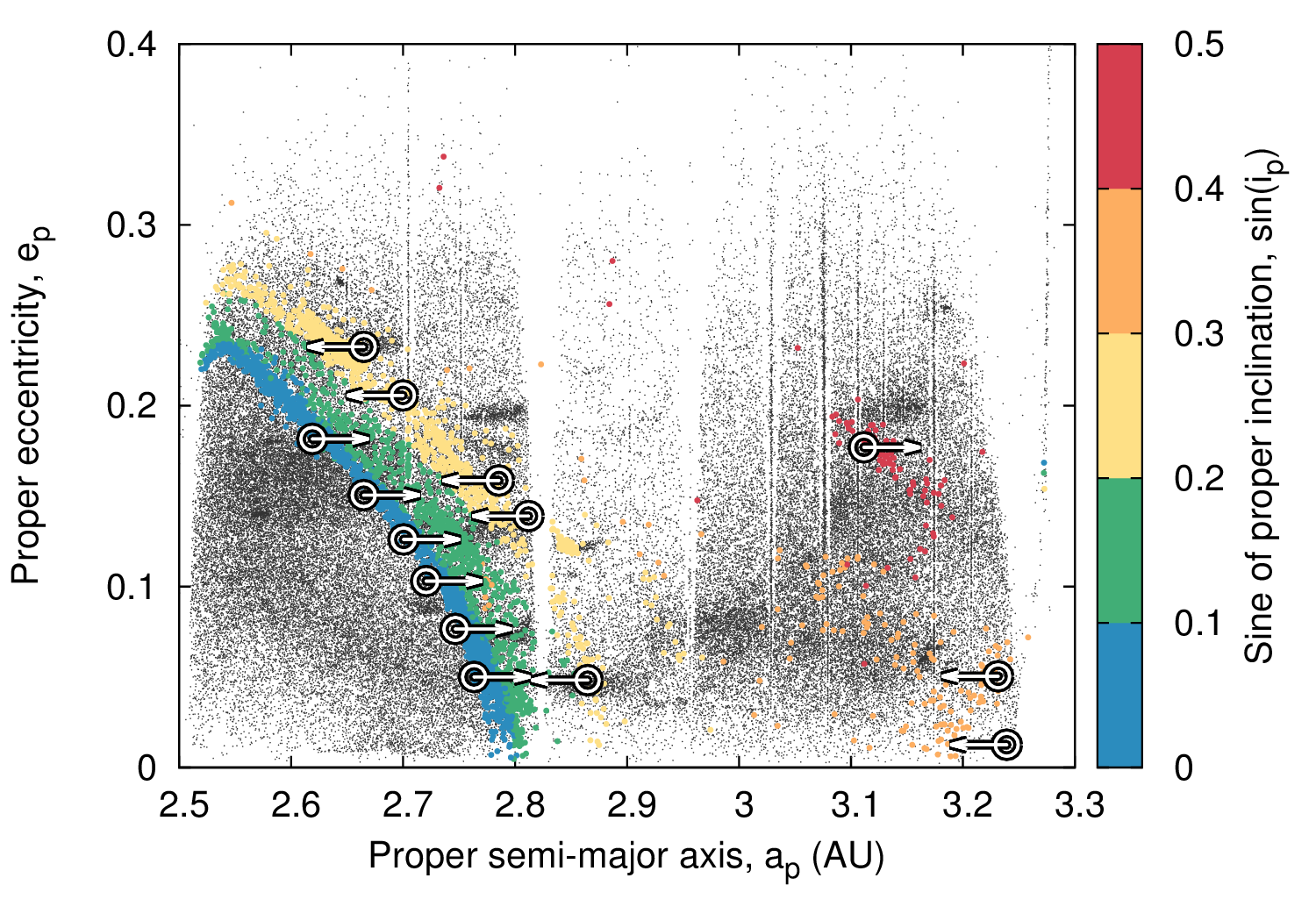} 
 \caption{The location of the groups of initial conditions for the simulations about the $v_{\rm {1c}}$ secular resonance on the $(a_{\rm p},e_{\rm p})$ plane. The gray dots represent all main belt asteroids, and the colored points the resonant ones for different inclinations according to the color code. Black circles denote the location and black arrows the Yarkovsky drift direction of each group of initial conditions.}
   \label{fig:v1cinit}
\end{center}
\end{figure}

In order to study the effect of the resonance, we considered three regions that are crossed by the resonance. Since as discussed above this secular resonance forms a strip like shape in the middle belt on the $(a_{\rm p},e_{\rm p})$ plane, it is intuitive to  choose the initial conditions for our test particles just outside this strip, so they are forced to cross the resonance by drifting in semi-major axis due to the Yarkovsky effect. This idea will also guide the selection of the initial conditions for the other secular resonances. Therefore, we can distinguish the relevant regions into the very low and moderate inclination parts of the middle belt, and the high inclination part of the outer belt. 

We created a number of groups of 20 test particles, as shown in \autoref{fig:v1cinit} for each region. The initial conditions of the particles within each group were generated with a variance of the order of $10^{-3}$ in the semi-major axis, eccentricity and inclination values, and identical random angular elements. We then we integrated  numerically their orbits within the two dynamical models we explained above. From the resulting time series of their mean orbital elements we calculated the evolution in time of their synthetic proper elements \citep{Knezevic2000} using running windows of $10\,{\rm {Myrs}}$ length, shifting them by $2\,{\rm {Myrs}}$ steps. As this secular resonance is a linear one involving only the proper frequency of the precession of the ascending node $(s)$ of the test particles, it only produces perturbations in their proper inclination and not in their eccentricity. Therefore we are only interested in the evolution of the proper inclination of the affected asteroids. The situation is the opposite for the secular resonances where the proper frequency of the precession of the longitude of perihelion $(g)$ is involved, perturbing only the eccentricities and not the inclinations of the asteroids.

\begin{figure}[h!]
\begin{center}
\includegraphics[width=\columnwidth]{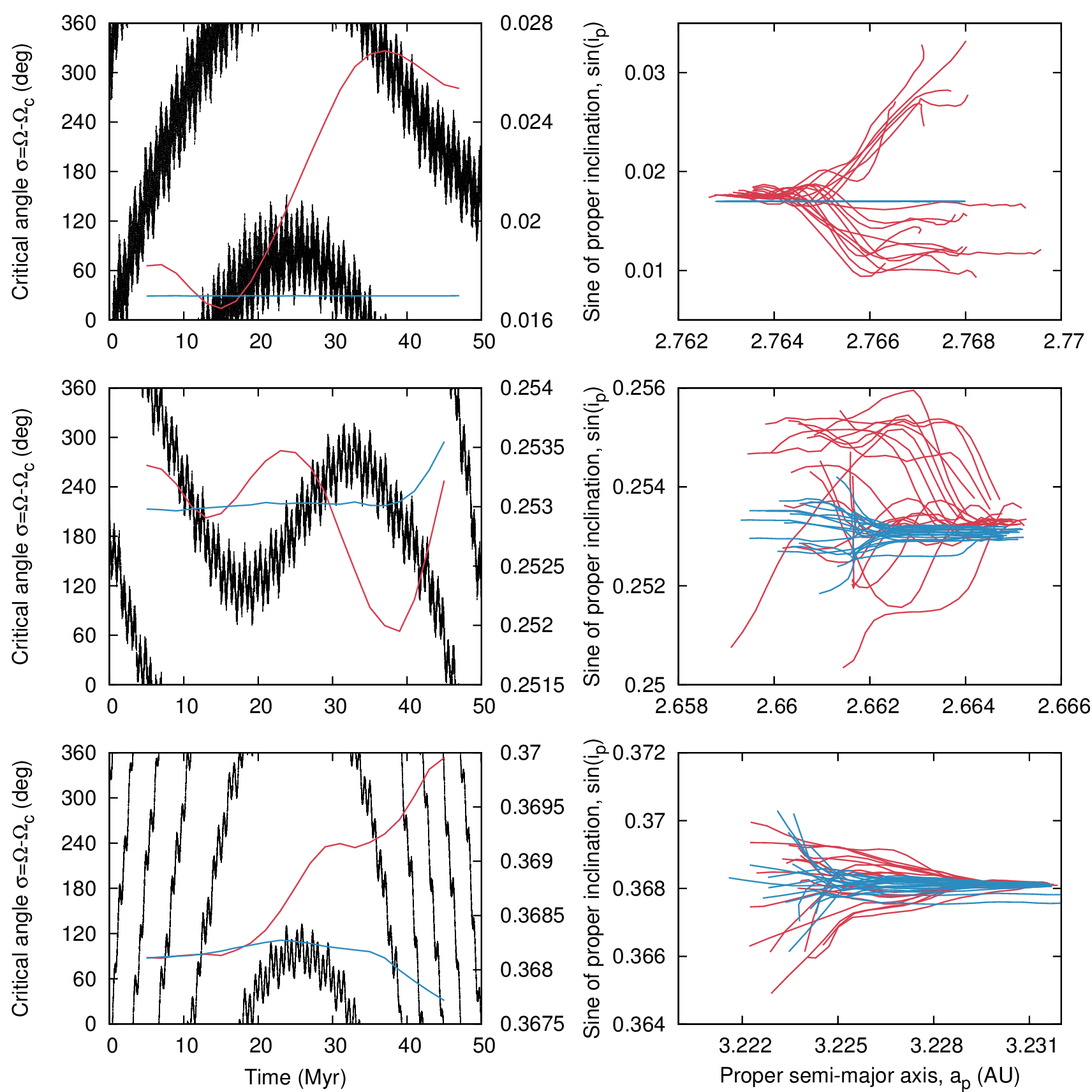} 
 \caption{Orbital evolution due to the secular resonance $\nu_{\rm {1c}}$ for the three representative regions. Top: low inclination middle belt, Mid: high inclination middle belt, Bottom: high inclination outer belt. Left panels: In black the evolution of the critical angle $\sigma=\Omega-\Omega_{\rm c}$ of a test particle.The red line shows the evolution of the sine of proper inclination of the same test particle with Ceres included in the model. The blue line shows the evolution of the proper inclination of the same test particle without Ceres in the model. Right panels: The evolution of the 20 test particles of the whole group in the two dynamical models, red with Ceres and blue without. }
   \label{fig:example}
\end{center}
\end{figure}

A representative example of the results for each region is shown in \autoref{fig:example}. The left panels show the evolution in time of the proper inclination of a single particle belonging to a group of initial conditions, plotted over the time evolution of the resonant critical angle $\sigma=\Omega-\Omega_{\rm c}$. We see that the crossing of the resonance, corresponding to the libration of the critical angle, results in excitation of the proper inclination when Ceres is included in the model as a perturber, whereas for the same initial condition the inclination of the orbit remains stable if we do not include Ceres. The right panels show the evolution in the proper semi-major axis versus sine of proper inclination plane $(a_{\rm p},\sin{i_{\rm p}})$ of the 20 particles of each group, in the two dynamical models. 

In \citet{Novakovic2015a} we have shown that asteroids entering the resonance experience oscillations in their inclination for as long as their critical angle librates, as seen in the left panel of \autoref{fig:example}. In order to quantify the effect of the perturbation induced by Ceres through the secular resonance, we measure the maximal change in the proper inclination of the test particles, as they cross the resonance.  For the low inclination middle belt we have measured an average amplitude of variations of the order of $\Delta\sin{i_{\rm p}}=7\cdot10^{-4}$  for the groups of test particles with semi-major axes close to that of Ceres $(a_{\rm {p(Ceres)}}=2.767\,{\rm {AU}})$, decreasing to $4\cdot10^{-4}$ as we move to lower semi-major axes towards $2.6\,{\rm {AU}}$ for our innermost group. For the high inclination middle belt we found an average amplitude of $3\cdot10^{-4}$. In the farther part of the outer belt ($a_{\rm p}>3\,{\rm {AU}}$), the amplitude of the oscillations is substantially smaller, around $1-2\cdot10^{-4}$, making it more difficult to separate the effect of the secular resonance from the other perturbing mechanisms that act in this region.

\subsubsection{The $\nu_{\rm {c}}$ resonance}
\begin{figure}[h!]
\begin{center}
\includegraphics[width=\columnwidth]{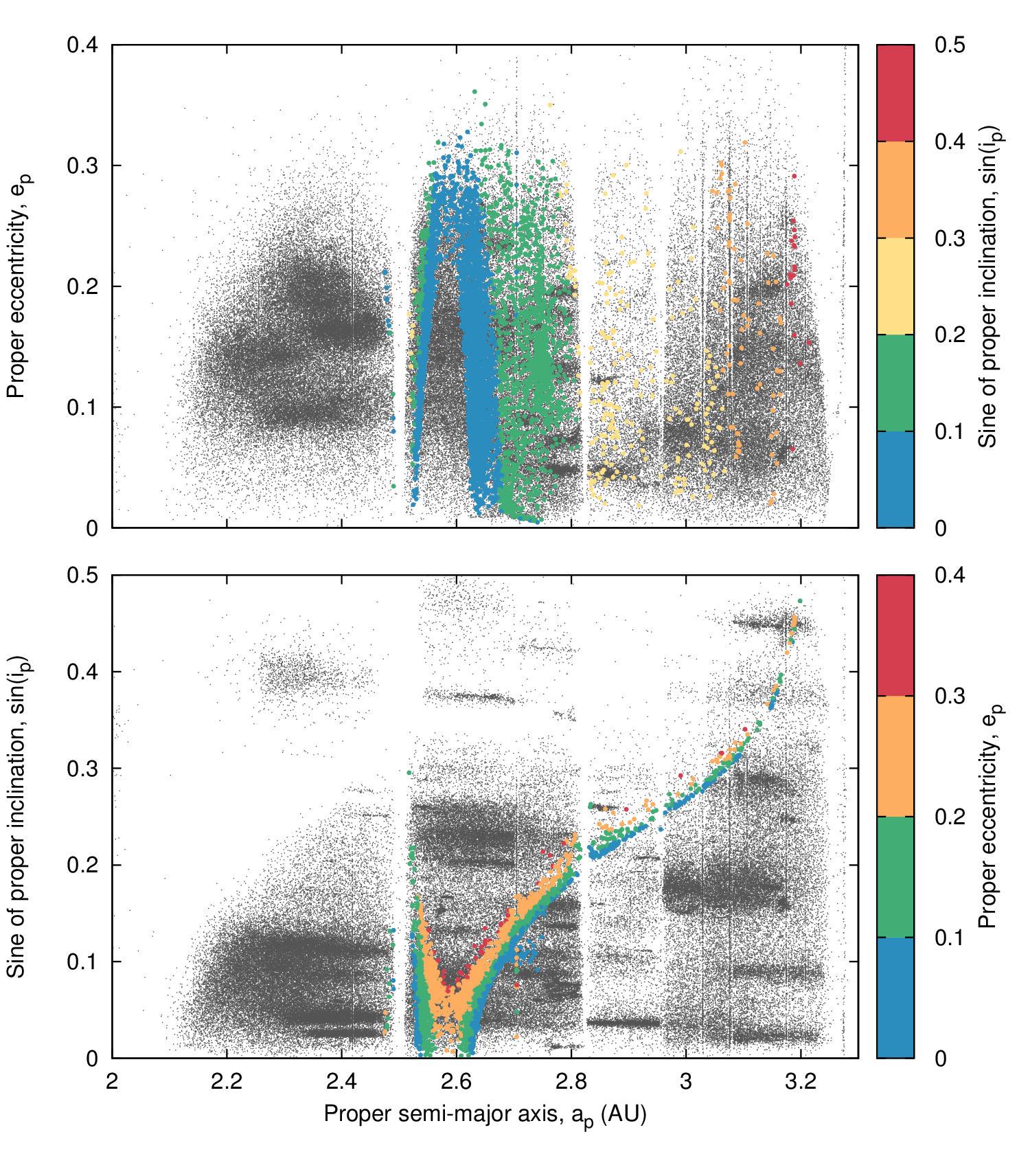} 
 \caption{The location of the $v_{\rm c}$ secular resonance on the $(a_{\rm p},e_{\rm p})$ plane (top), and on the $(a_{\rm p},\sin{i_{\rm p}})$ plane (bottom). The gray dots represent all main belt asteroids, and the colored points the resonant ones for different inclinations (top panel) and eccentricities (bottom panel) according to the respective color codes given in the legend.}
   \label{fig:vc}
\end{center}
\end{figure}

Following the same procedure we find the location of the $\nu_{\rm {c}}$ secular resonance, by plotting the asteroids that satisfy the resonant relation $|g-g_{\rm c}|\leq0.2''/{\rm {yr}}$. The result is shown in \autoref{fig:vc}, showing the proper semi-major axis versus the sine of proper inclination of the main belt $(a_{\rm p},\sin{i_{\rm p}})$, with the resonant asteroids highlighted in color for different proper eccentricities. This secular resonance also crosses mostly the middle part of the main belt, as well as the high inclination part of the outer belt.

\begin{figure}[h!]
\begin{center}
\includegraphics[width=\columnwidth]{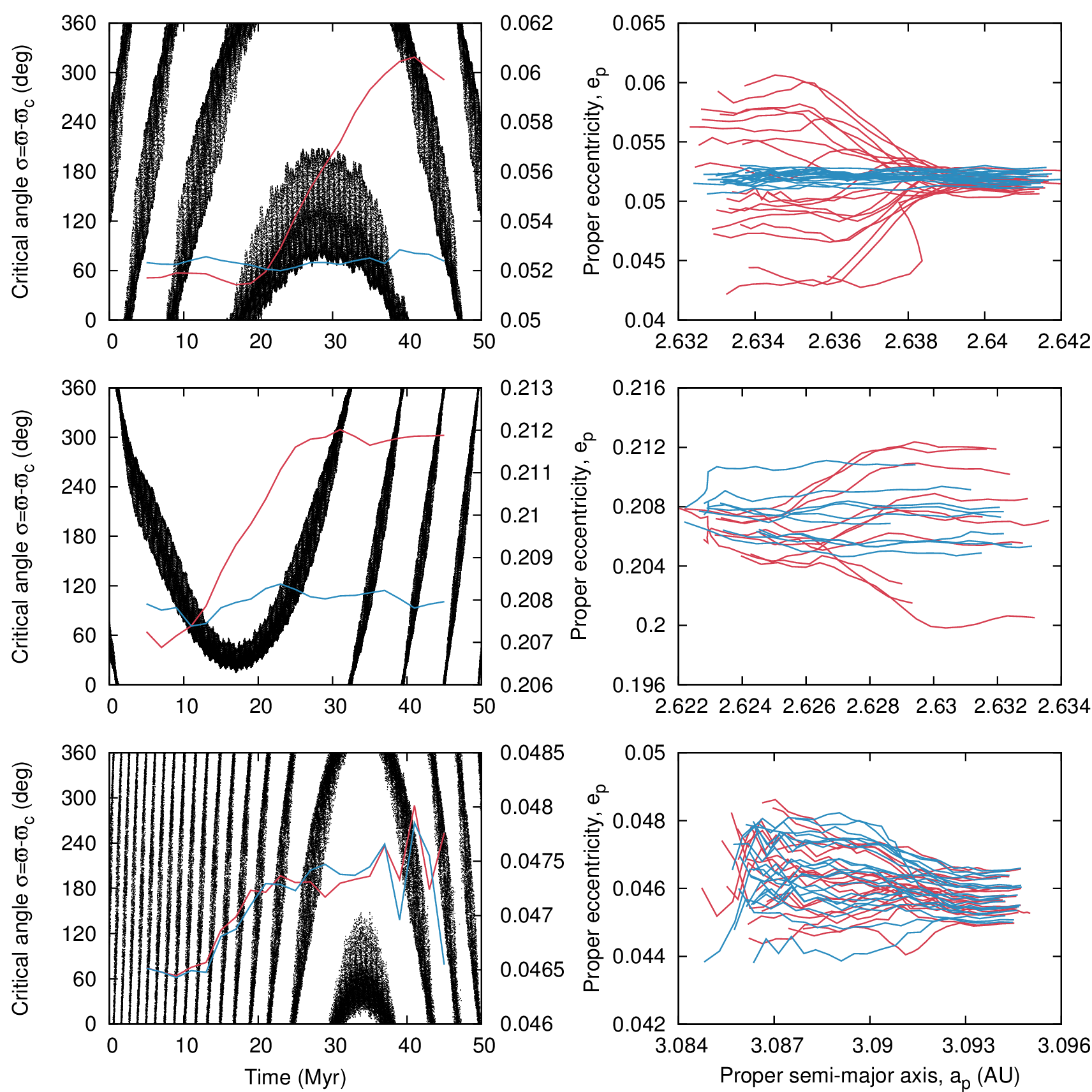} 
 \caption{Orbital evolution due to the secular resonance $\nu_{\rm c}$ for the three representative regions. Top: low eccentricity middle belt, Mid: high eccentricity middle belt, Bottom: outer belt. Left panels: In black the evolution of the critical angle $\sigma=\varpi-\varpi_{\rm c}$ of a test particle.The red lines show the evolution of the proper eccentricity of the same test particle with Ceres included in the model. The blue lines show the evolution of the proper eccentricity of the same test particle without Ceres in the model. Right panels: The evolution of the 20 test particles of the whole group in the two dynamical models, red with Ceres and blue without. }
   \label{fig:examplevc}
\end{center}
\end{figure}

Continuing our previous approach, we distinguish three regions to focus our study on: The low eccentricity $(e_{\rm p}<0.05)$ and high eccentricity $(e_{\rm p}>0.2)$ parts of the middle belt, and the outer belt. Representative examples of the behavior of asteroid orbits in these three regions, resulting from our numerical integrations of test particles are shown in \autoref{fig:examplevc}. We determine the strength of this resonance by the amplitude of the induced oscillations in proper eccentricity, as shown in the example of \autoref{fig:examplevc}. For the low eccentricity middle belt we found a maximum amplitude of $0.01$, for test particles close to Ceres (in terms of semi-major axis), decreasing to $0.003$ as we move further away. For the high eccentricity middle belt and inner part of the outer belt, the measured amplitude was of the order of $0.003$, while the relative strength of the perturbations from other causes increased significantly. For the farther part of the outer belt ($a_{\rm p}>3\,{\rm {AU}}$), although we have a clear signature from the critical angle that the test particles cross the secular resonance, its impact on the eccentricities of the orbits is effectively zero, as the two models give statistically indistinguishable results, as can be seen in the bottom part of \autoref{fig:examplevc}     

\subsection{Secular resonances with Vesta}
\subsubsection{The $\nu_{\rm {1v}}$ resonance}

\begin{figure}[h!]
\begin{center}
\includegraphics[width=\columnwidth]{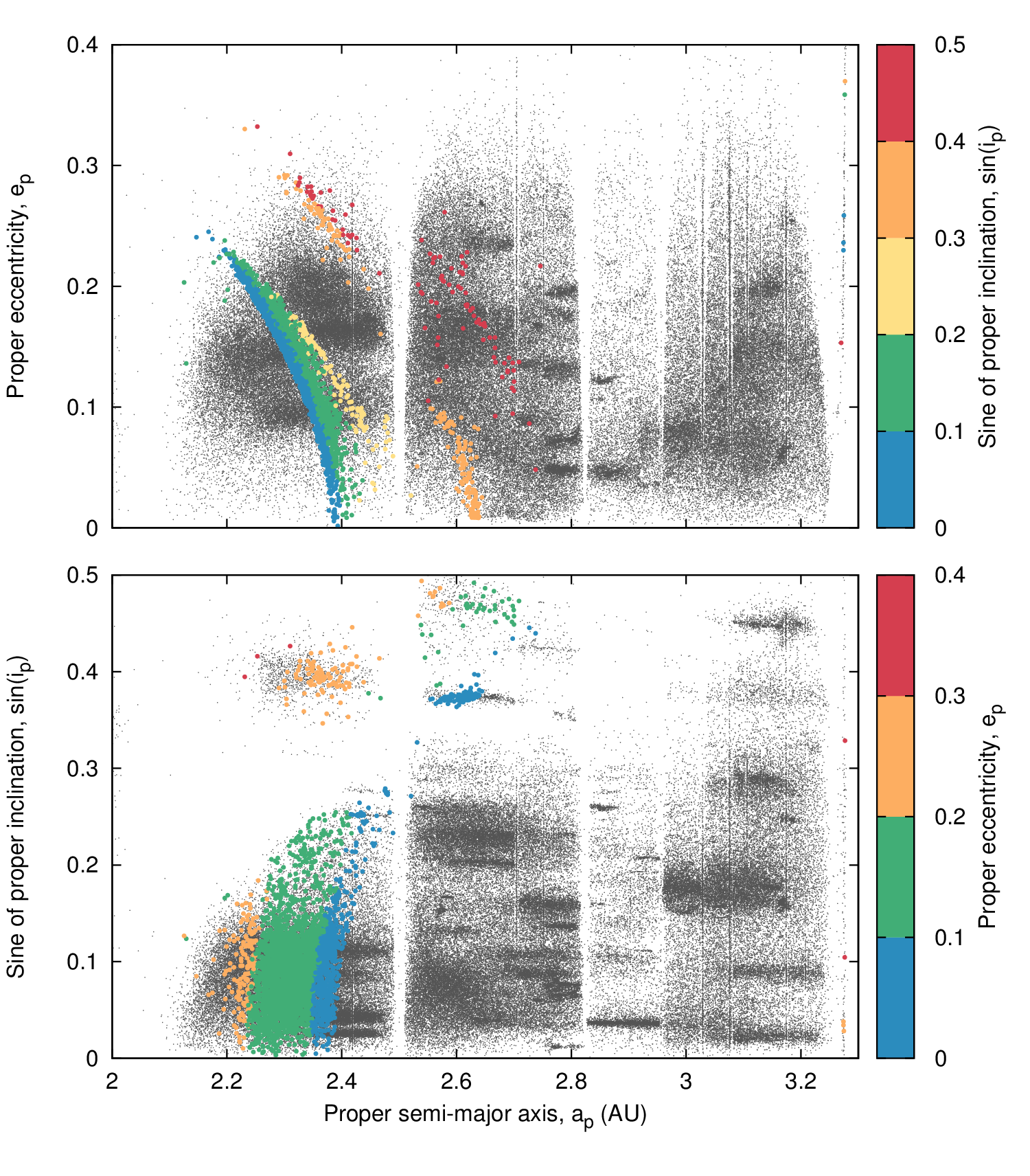} 
 \caption{The location of the $v_{\rm {1v}}$ secular resonance on the $(a_{\rm p},e_{\rm p})$ plane (top) and on the $(a_{\rm p},\sin{i_{\rm p}})$ plane (bottom). The gray dots represent all main belt asteroids, and the colored points the resonant ones for different inclinations (top panel) and eccentricities (bottom panel) according to the respective color codes given in the legend.}
   \label{fig:v1v}
\end{center}
\end{figure}

As the asteroid (4)~Vesta is located in the inner $(2<a_{\rm p}<2.5\,{\rm {AU}})$main belt, we expect the secular resonances involving it to predominantly affect this region. Indeed in \autoref{fig:v1v} we see the location of the $\nu_{\rm {1v}}$ secular resonance  by highlighting the asteroids with proper frequencies that satisfy the relation $|s-s_{\rm v}|=0.2''/{\rm {yr}}$, where we see that the inner belt is crossed by the resonance in a wide range of eccentricities and inclinations, while there are also some resonant asteroids with high inclinations in the middle belt.

\begin{figure}[h!]
\begin{center}
\includegraphics[width=\columnwidth]{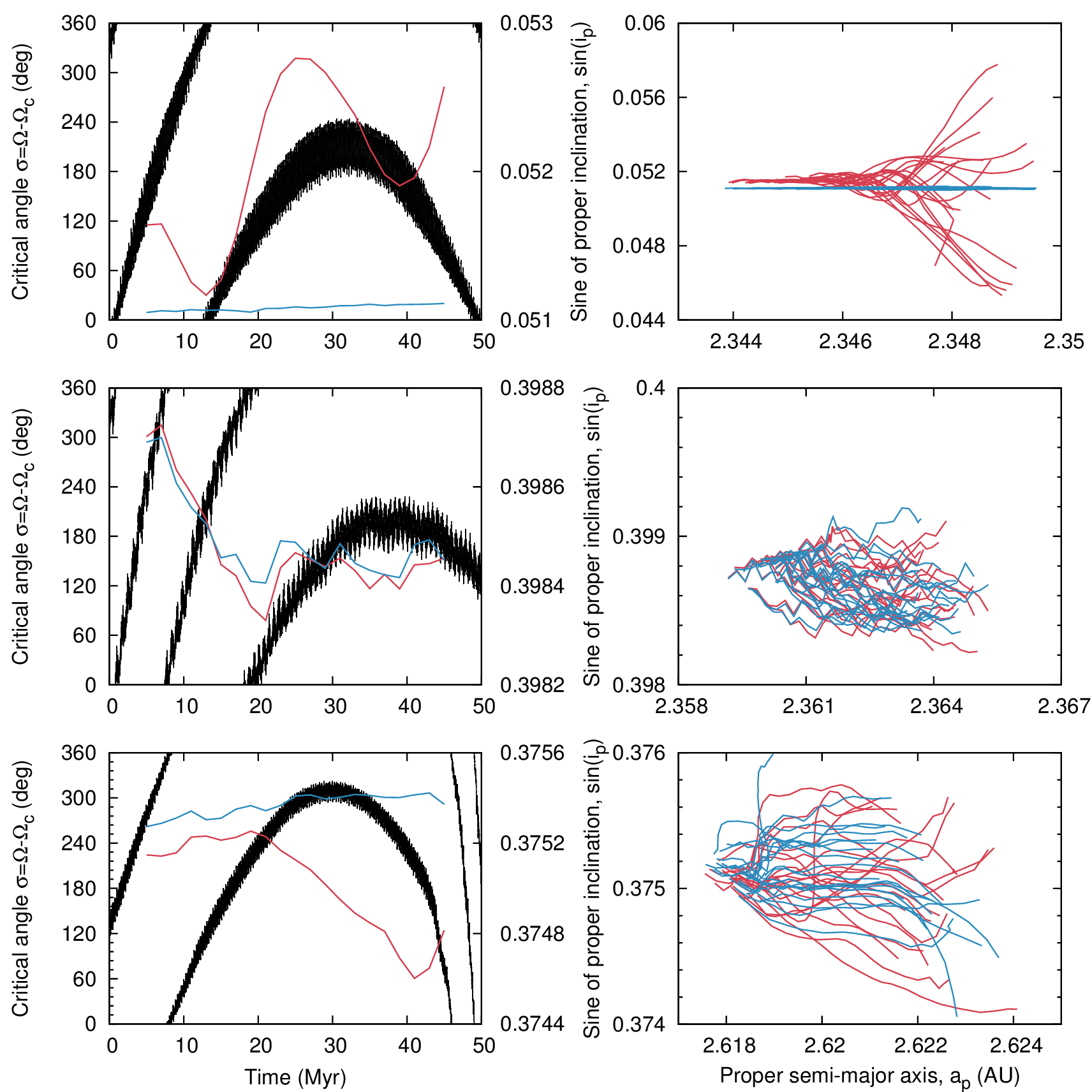} 
 \caption{Orbital evolution due to the secular resonance $\nu_{\rm {1v}}$ for the three representative regions. Top: low inclination inner belt, Mid: high inclination inner belt, Bottom: high inclination middle belt. Left panels: In black the evolution of the critical angle $\sigma=\Omega-\Omega_{\rm v}$ of a test particle.The red line shows the evolution of the sine of proper inclination of the same test particle with Vesta included in the model. The blue line shows the evolution of the proper inclination of the same test particle without Vesta in the model. Right panels: The evolution of the 20 test particles of the whole group in the two dynamical models, red with Vesta and blue without.}
   \label{fig:examplev1v}
\end{center}
\end{figure}

The situation with this resonance is slightly different than with the ones involving Ceres. When we examine the path of the resonance in \autoref{fig:v1v}, we see that the high inclination region of the inner belt is also highly eccentric, while the high inclination resonant region of the middle belt has also a low eccentricity part. This led to the result we present in \autoref{fig:examplev1v}, that is the high inclination part of the inner belt, despite being close in semi-major axis to Vesta, shows no distinctive evolution caused by the resonance, whereas the resonant region in the middle belt, has a very small $(\sim0.0002)$, but identifiable signature of inclination excitation due to the resonance. The low inclination part of the inner belt is showing as expected the largest amplitudes of oscillations in the sine of inclination, of the order of $0.004$. 

\subsubsection{The $\nu_{\rm {v}}$ secular resonance}

\begin{figure}[h!]
\begin{center}
\includegraphics[width=\columnwidth]{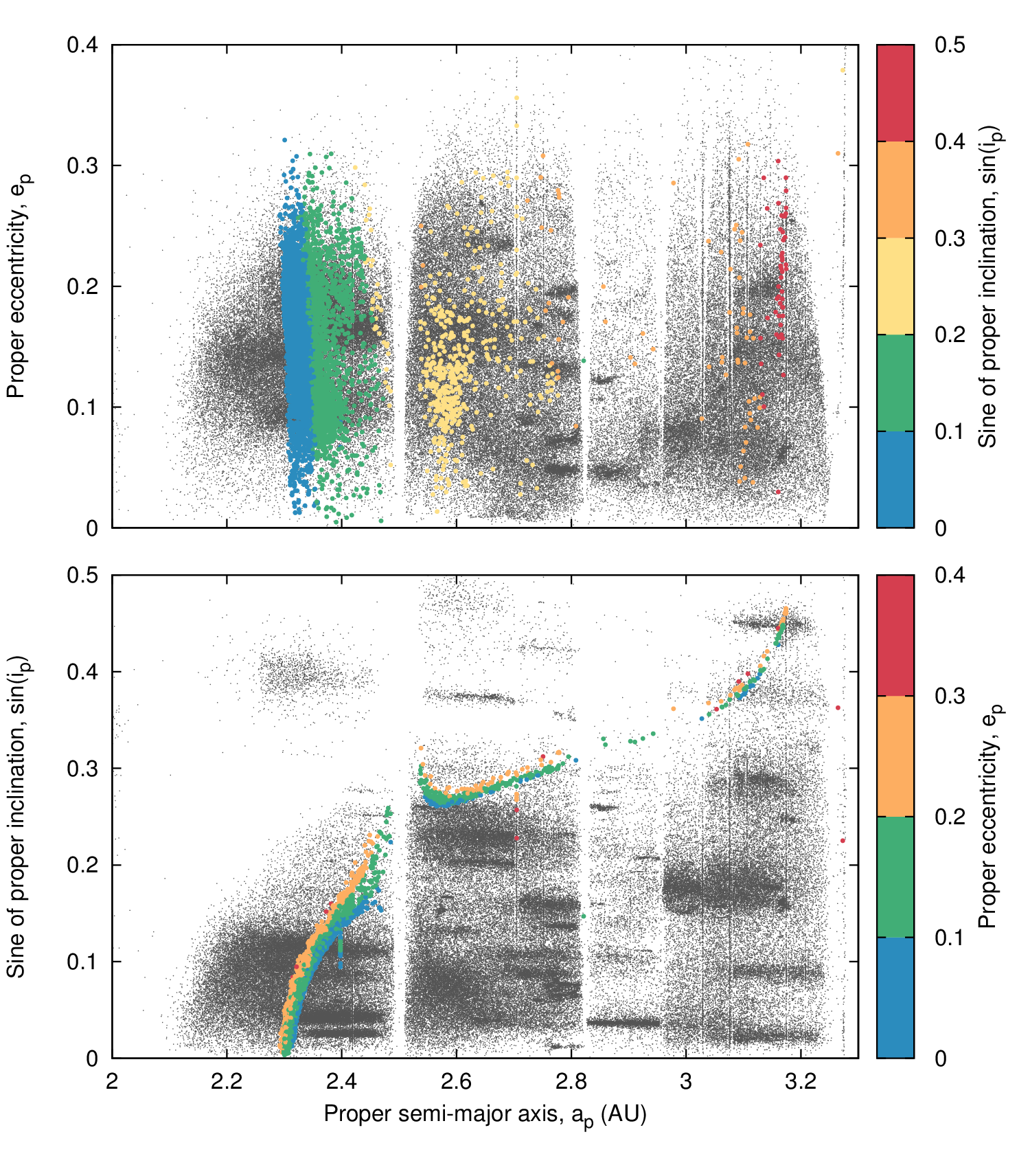} 
 \caption{The location of the $v_{v}$ secular resonance on the $(a_{\rm p},e_{\rm p})$ plane (top), and on the $(a_{\rm p},\sin{i_{\rm p}})$ plane (bottom). The gray dots represent all main belt asteroids, and the colored points the resonant ones for different inclinations (top panel) and eccentricities (bottom panel) according to the respective color codes given in the legend.}
   \label{fig:vv}
\end{center}
\end{figure}

\begin{figure}[h!]
\begin{center}
\includegraphics[width=\columnwidth]{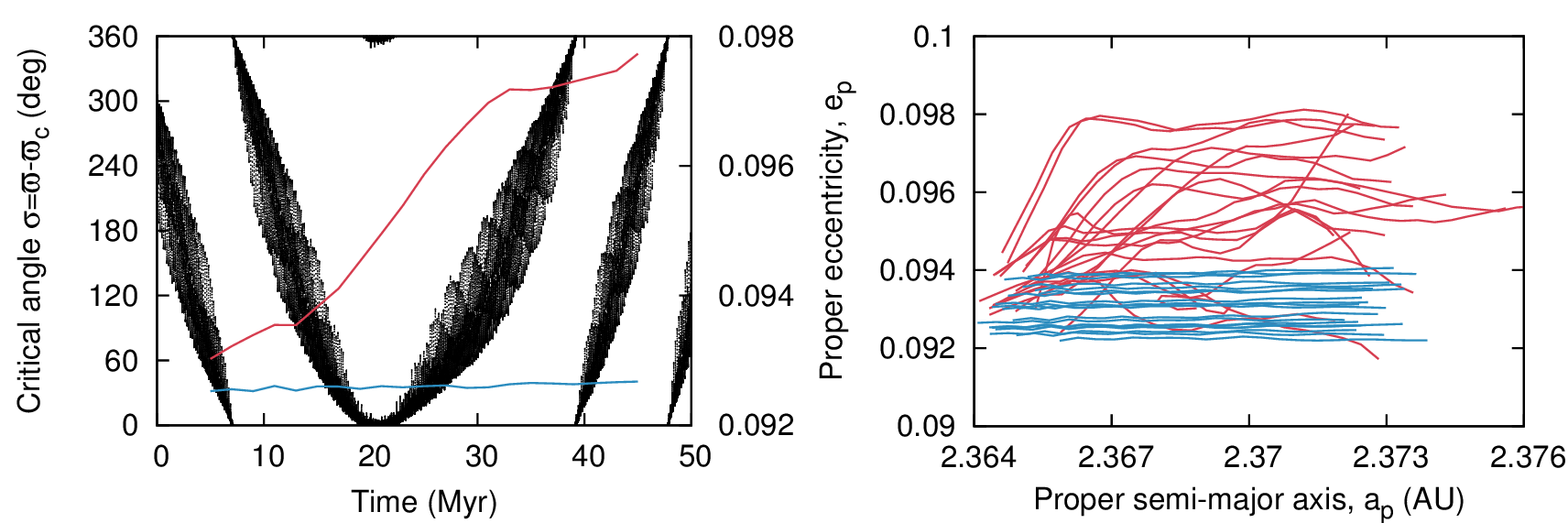} 
 \caption{Orbital evolution due to the secular resonance $\nu_{\rm {v}}$ for the inner belt. Left panel: In black the evolution of the critical angle $\sigma=\Omega-\Omega_{\rm v}$ of a test particle.The red line shows the evolution of the proper eccentricity of the same test particle with Vesta included in the model. The blue line shows the evolution of the proper eccentricity of the same test particle without Vesta in the model. Right panel: The evolution of the 20 test particles of the whole group in the two dynamical models, red with Vesta and blue without.}
   \label{fig:exvv}
\end{center}
\end{figure}

The last secular resonance we studied is the one involving the precession frequency of the perihelion of (4)~Vesta, namely $\nu_{\rm {v}}$. \autoref{fig:vv} shows the location of the asteroids whose proper frequencies $g$ satisfy the relation $|g-g_{\rm v}|\leq0.2''/{\rm {yr}}$, revealing the location of the secular resonance across the main belt as in the previous cases. In the $(a_{\rm p},\sin{i_{\rm p}})$ plane we notice the pretty clear path of the resonance, crossing the inner belt from low to moderate inclinations, continuing to the high inclination part of the middle belt and on to a very high inclination range of the outer main belt, always covering a very wide range of eccentricities as can be seen in the $(a_{\rm p},e_{\rm p})$ plane. For this resonance we focused our numerical simulations on the inner belt only. The method we used for revealing the effect of each resonance depends on the action of the Yarkovsky effect in order to force the test particles through the secular resonances. This means that it is difficult to apply this scheme if a secular resonance's path is parallel, or almost parallel, to the $a_{\rm p}$ axis, as is the case for the $\nu_{\rm {v}}$ secular resonance in the middle belt, and for this reason we did not manage to investigate this part. In the inner belt we found oscillations in proper eccentricity with amplitudes of the order of $0.004$ as shown in \autoref{fig:exvv}.

The results for all the cases we investigated are summarized in \autoref{Table:1}. Where ranges are given, the largest value corresponds to asteroids with proper semi-major axes close to those of the respective perturbing body (Ceres or Vesta). We notice that the maximal values of the changes in proper inclination and eccentricity caused by Ceres are almost two times bigger compared to the ones caused by Vesta, a consequence of the fact that Ceres is approximately 3.5 times more massive than Vesta, thus exerting stronger perturbations as expected.  

\begin{table}[]
\centering
\caption{Summary table of the maximal changes in the proper elements of the main belt asteroids caused by the secular resonances with Ceres and Vesta.}
\label{Table:1}
\resizebox{\columnwidth}{!}{
\begin{tabular}{|l|l|c|c|c|}
\hline
\multicolumn{1}{|c|}{\multirow{2}{*}{Secular resonance}} & \multirow{2}{*}{Measured quantity} & \multicolumn{3}{c|}{Range in $a_{\rm p}\,({\rm {AU}})$}              \\ \cline{3-5} 
\multicolumn{1}{|c|}{}                                   &                                    & $2<a_{\rm p}<2.5$     & $2.5<a_{\rm p}<3$                   & $a_{\rm p}>3$         \\ \hline
$\nu_{\rm {1c}}$                                               & $\Delta\sin(i_{\rm p})$                  & -               & $4-7\cdot10^{-4}$             & $2\cdot10^{-4}$ \\ \hline
$\nu_{\rm c}$                                                & $\Delta e_{\rm p}$                       & -               & $3\cdot10^{-3}-1\cdot10^{-2}$ & $3\cdot10^{-3}$ \\ \hline
$\nu_{\rm {1v}}$                                               & $\Delta\sin(i_{\rm p})$                  & $4\cdot10^{-4}$ & $2\cdot10^{-4}$               & -               \\ \hline
$\nu_{\rm {v}}$                                                & $\Delta e_{\rm p}$                       & $4\cdot10^{-3}$ & -                             & -               \\ \hline
\end{tabular}
}
\end{table}

\subsection{Asteroid families}
One important aspect of the action of the secular resonances with massive asteroids we have presented is the effect they may have on the orbital evolution of asteroid family members. Since the asteroid families are more or less compact in the space of proper elements, the action of the secular resonances should give a distinct signature, identifiable merely by the shape of the family member distributions in the different projections of the proper elements. Indeed in \citet{Novakovic2015a} we have shown that the asymemtric shape in the proper semi-major axis versus proper inclination plane $(a_{\rm p},\sin{i_{\rm p}})$ of the Hoffmeister family is caused by the $v_{\rm {1c}}$ secular resonance with Ceres. Also in \citet{Novakovic2016} we have shown that this case is not unique, as the asteroid families (1128) Astrid and (1521) Seinajoki, also owe their irregular shapes in the proper elements space to the $v_{\rm {1c}}$ secular resonance.

Although it is out of the scope of this work to study individual asteroid families for possible interactions with the secular resonances, as our aim is a more global view of their importance, we find it worthy to present which asteroid families are expected to be influenced. This is done in a similar way as our numerical method of finding the location of the resonances. Instead of looking at the whole catalog of proper elements for resonant asteroids, we are instead looking in the catalog of only those asteroids that belong to asteroid families. For this we use the classification of \citet{Milani2014}. In this way we can find which families are crossed by the secular resonances we present here and which, if any, show signs of interaction with them.

\subsubsection{Asteroid families interacting with the $\nu_{\rm {1c}}$ secular resonance} 
\begin{figure}[h!]
\begin{center}
\includegraphics[width=\columnwidth]{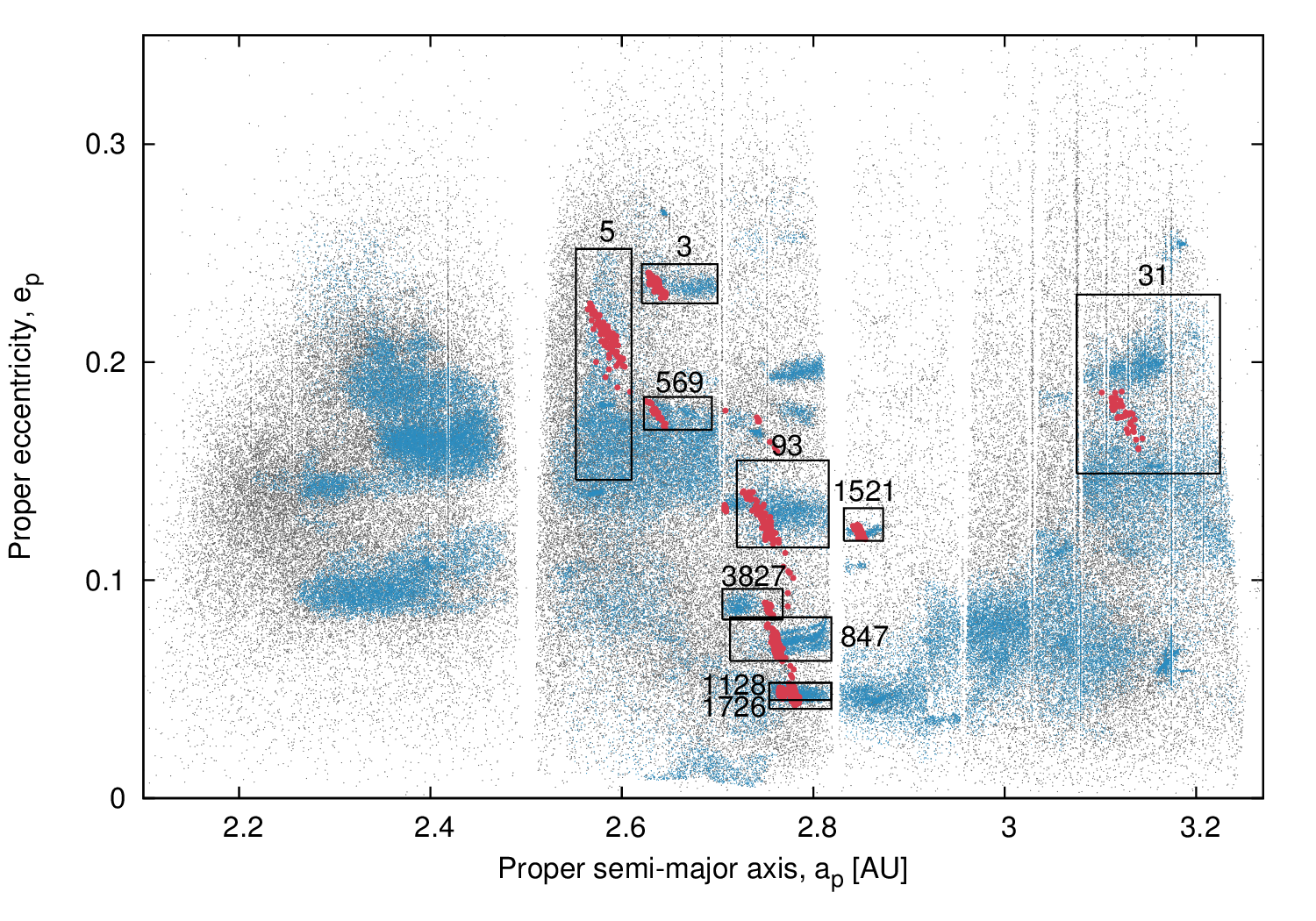} 
 \caption{Asteroid families crossed by the secular resonance $\nu_{\rm {1c}}$ with Ceres. Gray dots represent all main melt asteroids, and blue dots those who belong to asteroid families. The red points represent resonant asteroids belonging to asteroid families (highlighted in black boxes).}
   \label{fig:v1cfam}
\end{center}
\end{figure}

Using the method described above, we find that ten asteroid families have a significant number of their members currently in resonance\footnote{By significant we mean a number of the order of at least ten asteroids in regular, non-chaotic orbits. We make this discrimination as there may be asteroids with proper frequencies that satisfy the resonant relation, but the error in their frequency is large, resulting from other effects such as a mean motion resonance.} as shown in \autoref{fig:v1cfam}. These families are: (3) Juno, (5) Astraea, (31) Euphrosyne, (93) Minerva, (569) Misa, (847) Agnia, (1128) Astrid, (1521) Seinajoki, (1726) Hoffmeister and (3827) Zdenekhorsky. 

Apart from the families of (1128) Astrid, (1521) Seinajoki and (1726) Hoffmeister which we have already studied separately, as mentioned above, the $\nu_{\rm {1c}}$ secular resonance may be of some importance for the families of (569) Misa (847) Agnia and (3827) Zdenenkovsky as these are close to (1)~Ceres in terms of semi-major axis, and cover ranges in the sine of proper inclination comparable to the magnitude of the induced perturbations as we measured them. 

The case of (847) Agnia may be of particular interest, as this family is also crossed by the $z_1=g+s-g_6-s_6$ secular resonance. Indeed in the $(a_{\rm p},\sin{i_{\rm p}})$ plane the two resonances cross the family in a perpendicular way with respect to each other, and because of that we discovered some hints that the secular resonance with Ceres might be able to drive asteroids out of the $z_1$. Of course, this requires further investigation to be proven, that is out of the scope of this work. 

The family of (31) Euphrosyne is another example of potential interaction between resonances, as it is crossed by a multitude of them. The secular resonances with the giant planets are more powerful than $\nu_{\rm {1c}}$ in this region, and play an important role in the evolution of the family \citep{Carruba2014}. Still it is possible that even a weak perturbation by $\nu_{\rm {1c}}$ may have an amplified effect due to the interaction with them.

Finally the family of (93) Minerva is crossed by the 3J-1S-1A three body resonance \citep{Nesvorny1998} at the same location where the $\nu_{\rm {1c}}$ crosses it, making the effect of the latter practically indistinguishable.

\subsubsection{Asteroid families interacting with the $\nu_{\rm {c}}$ secular resonance}
\begin{figure}[h!]
\begin{center}
\includegraphics[width=\columnwidth]{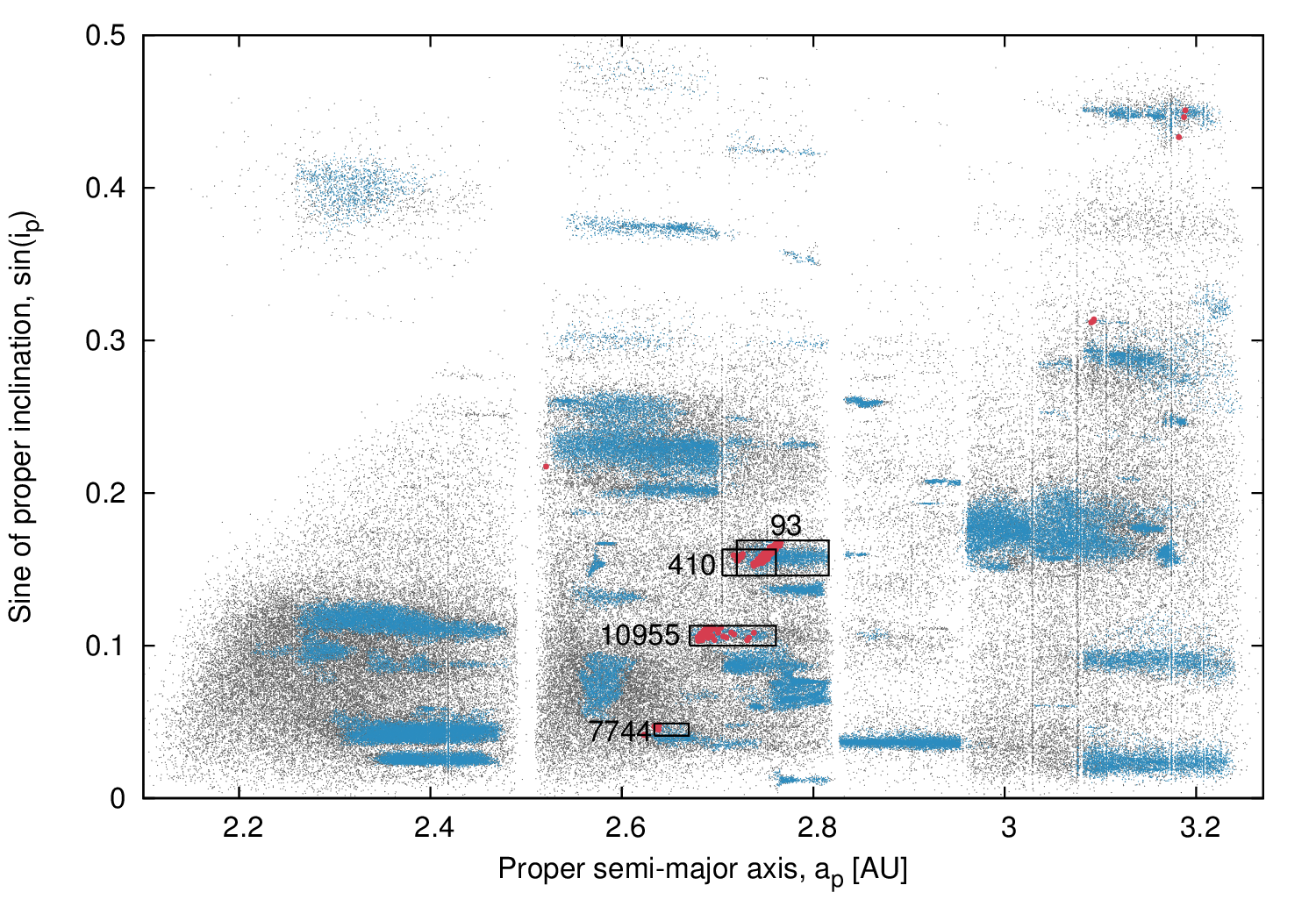} 
 \caption{Asteroid families crossed by the secular resonance $\nu_{\rm c}$ with Ceres. Gray dots represent all main melt asteroids, and blue dots those who belong to asteroid families. The red points represent resonant asteroids belonging to asteroid families (highlighted in black boxes).}
   \label{fig:vcfam}
\end{center}
\end{figure}

In the same manner we identify the asteroid families that are crossed by the $\nu_{\rm {c}}$ secular resonance, shown in \autoref{fig:vcfam}. The families crossed by this resonance are: (93) Minerva, (410) Chloris, (7744) 1986QA$_{1}$ and (10955) Harig. Of these families (410) Chloris and (7744) 1986QA$_{1}$ are narrow enough in proper eccentricity so that the secular resonance could be of some importance in their evolution whereas (93) Minerva and (10955) Harig might also seem to be good candidates for further study, as they are large families and their shapes suggest possible influence by the secular resonance. However such a study is not trivial as for the case of (93) Minerva the $\nu_{\rm {c}}$ secular resonance and the 3J-1S-1A overlap,  as in the previous case, and the latter dominates the perturbations in eccentricity, whereas Harig is in a place where many secular resonances with the giant planets converge, making it impossible to distinguish the effect of Ceres.

\subsubsection{Asteroid families interacting with the $\nu_{\rm {1v}}$ secular resonance}

\begin{figure}[h!]
\begin{center}
\includegraphics[width=\columnwidth]{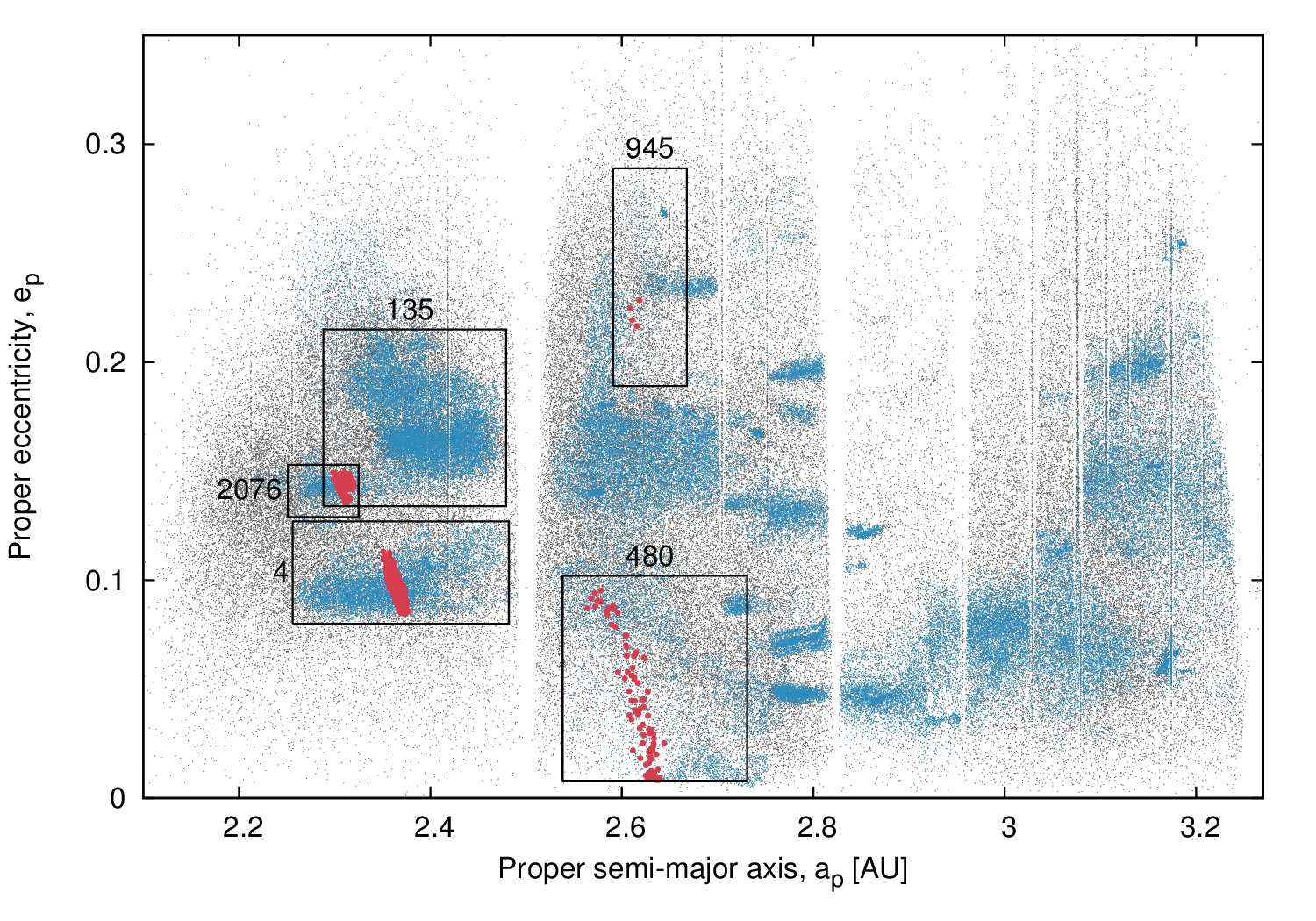} 
 \caption{Asteroid families crossed by the secular resonance $\nu_{\rm {1v}}$  with Vesta. Gray dots represent all main melt asteroids, and blue dots those who belong to asteroid families. The red points represent resonant asteroids belonging to asteroid families (highlighted in black boxes).}
   \label{fig:v1vfam}
\end{center}
\end{figure}

In \autoref{fig:v1vfam} we present the results for the case of the $\nu_{\rm {1v}}$ secular resonance with Vesta. We found five families that are crossed by the resonance, which are: (4)~Vesta, (135) Hertha, (480) Hansa, (945) Barcelona and (2076) Levin. Our interest for this case is drawn not in the big families, where nothing special seems to happen, but at the very high inclination family of (945) Barcelona. The size of this family in the proper elements space is comparable to the magnitude of the perturbations given by the $\nu_{\rm {1v}}$ secular resonance, and it shows some hints of irregular shape where it is crossed by it. Even the possibility that a secular resonance with Vesta might be important at such a high inclination in the middle belt is intriguing, and deserves further study.

\subsubsection{Asteroid families interacting with the $\nu_{\rm {v}}$ secular resonance}

\begin{figure}[h!]
\begin{center}
\includegraphics[width=\columnwidth]{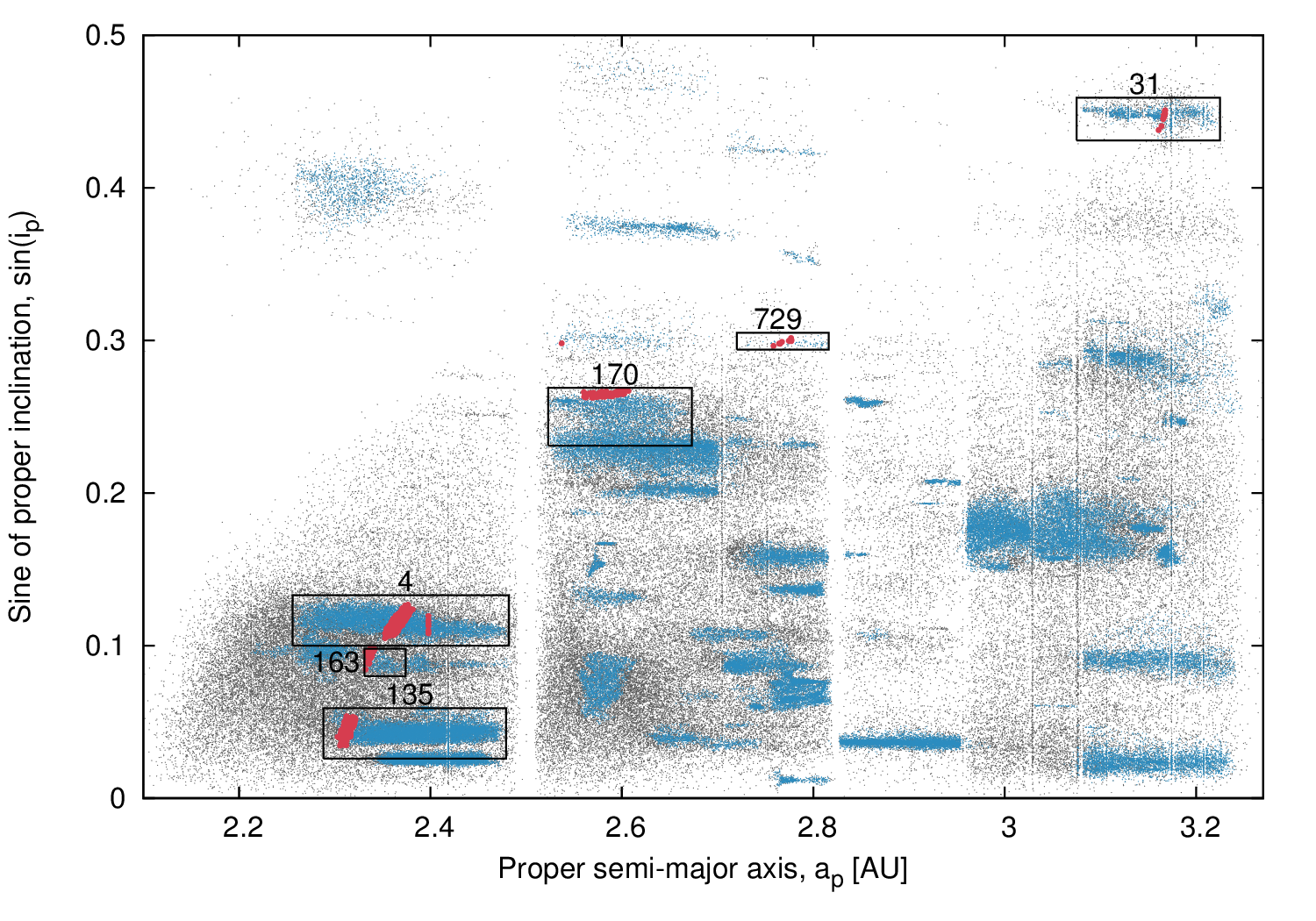} 
 \caption{Asteroid families crossed by the secular resonance $\nu_{\rm {v}}$  with Vesta. Gray dots represent all main melt asteroids, and blue dots those who belong to asteroid families. The red points represent resonant asteroids belonging to asteroid families (highlighted in black boxes).}
   \label{fig:vvfam}
\end{center}
\end{figure}

In \autoref{fig:vvfam} we present the asteroid families which we found to be crossed by the last secular resonance we consider here, the $\nu_{\rm {v}}$ secular resonance with Vesta. We found six such families, namely: (4)~Vesta, (31) Euphrosyne, (135) Hertha, (163) Erigone, (170) Maria and (729) Watsonia.  However, we were unable to relate any specific property of these families to the existence of the resonance, as these families are either too large, in which case the perturbations can not lead to significant alteration of their shape, or too far away from Vesta, where the perturbations are not strong enough.

In any case, the effect of the resonances on the shapes of the families
depends on whether the capture inside the resonance is long-lasting or
not,
as well as on the path of the resonance with respect to the family.
Therefore, a precise answer if the asteroidal secular resonances
play an important role in the dynamics of any of the families, requires
detailed study of each particular family. In this respect, the families
mentioned in this work should only be considered as good candidates for
this kind of investigation.

\section{Discussion and conclusions}

We have found the locations of the four linear secular resonances with (1)~Ceres and (4)~Vesta using a numerical approach that identifies asteroids which according to their proper frequencies appear to be in resonance. The secular resonances with Ceres mostly cover the middle part of the main belt, with some extension to the high inclination part of the outer belt, whereas those with Vesta cover the inner belt and a moderate to high inclination part of the middle  and outer belt.

Our numerical simulations have shown that the effects of these resonances on the orbits of main belt asteroids is considerable, especially when the latter have semi-major axes close to the respective perturbing massive asteroid. \citet{Milani1992,Milani1994} have studied the effect of non-linear secular resonances with the giant planets on the proper elements of main-belt asteroids. They found that resonant asteroids' proper elements undergo secular oscillations with amplitudes comparable to what we measured for the secular resonances with Ceres and Vesta.  In the outer belt, which is considered to be far enough from both, we cannot clearly distinguish the impact of the secular resonances among the other dynamical mechanisms that act in the region. Although, as we have shown, the effect of the latter diminishes with increasing distance from the relevant massive asteroid in each case, it is crucial to note that in specific regions of the main belt, secular resonances with massive asteroids are equally, if not more important than the non-linear ones with the giant planets. 

Finally we have identified which asteroid families are crossed by each resonance. There are cases where the size of the families in the proper elements space is comparable to the amplitude of the oscillations induced by the secular resonance that crosses them (e.g. 1726 Hoffmeister). In these cases the secular resonances studied here should have the most evident effect on the post-impact evolution of asteroid family members. 

\section*{Acknowledgments}

This work has been supported by the European Union [FP7/2007-2013], project: 
STARDUST-The Asteroid and Space Debris Network. BN also acknowledges support 
by the Ministry of Education, Science and Technological Development of the Republic 
of Serbia, Project 176011. Numerical simulations were run on the PARADOX-III cluster
hosted by the Scientific Computing Laboratory of the Institute of Physics Belgrade.
The authors would like to thank the referees Miroslav Bro{\v z} and Valerio Carruba for their constructive comments which helped improve the quality of this article.

\bibliography{bib}
\end{document}